# Field misalignment effect on the in-plane magnetoresistance in the high-temperature superconductor $FeSe/SrTiO_3$

Jie Liu,[1,*] Zihan Cui,[1,*] Xiaorong Liu,[1] Yanzhao Liu,[2] Ziqiao Wang,[2,†] Jian Wang,[3,4,5,6,‡] and Yi Liu,[1,2,7,§]

[1] *School of Physics and Key Laboratory of Quantum State Construction and Manipulation (Ministry of Education), Renmin University of China, Beijing 100872, China*
[2] *Quantum Science Center of Guangdong–Hong Kong–Macao Greater Bay Area (Guangdong), Shenzhen 518045, China*
[3] *International Center for Quantum Materials, School of Physics, Peking University, Beijing 100871, China*
[4] *Beijing Key Laboratory of Quantum Devices, Peking University, Beijing 100871, China*
[5] *Hefei National Laboratory, Hefei 230088, China*
[6] *Collaborative Innovation Center of Quantum Matter, Beijing 100871, China*
[7] *Xiongan Future Industrial Institute, Renmin University of China, Xiongan 071700, China*

[*] These authors contributed equally to this work.
[†] Contact author: wangziqiao@quantumsc.cn
[‡] Contact author: jianwangphysics@pku.edu.cn
[§] Contact author: yiliu@ruc.edu.cn

The azimuthal angular dependence of the in-plane magnetoresistance provides a powerful probe for detecting rotational symmetry breaking in superconductors, which has significant implications for understanding the superconducting pairing symmetry and the intertwined orders. However, in two-dimensional or layered superconductors with strong anisotropy between in-plane and out-of-plane critical fields, measuring the in-plane magnetoresistance is particularly challenging as it is extremely sensitive to the field misalignment. Here we perform systematic electrical transport measurements on ultrathin FeSe films grown on SrTiO3 substrates. By remounting the same FeSe film to change the field misalignment, two qualitatively different angular dependences of the in-plane magnetoresistance are observed in the superconducting transition region, indicating the extreme sensitivity of the measurement to field misalignment. The observed anisotropic in-plane magnetoresistance can be well captured by our geometric model considering the field misalignment effect. Our work provides a practical framework for disentangling the misalignment effect and investigating the in-plane rotational symmetry breaking in two-dimensional or layered superconductors.

## INTRODUCTION

Rotational symmetry breaking is a central topic in emergent quantum states and provides critical insights into unconventional superconductivity and intertwined orders [1–8]. The angular dependent in-plane magnetoresistance $R(\varphi)$ measurements serve as a direct probe of rotational symmetry breaking, and have been extensively utilized in a broad range of correlated electronic systems, including high-temperature superconductors [7,9,10], kagome superconductors [11] and $KTaO_3$ heterointerface superconductors [12]. However, for two-dimensional (2D) or layered superconductors with highly anisotropic upper critical fields, even a slight field misalignment relative to the sample plane can introduce an out-of-plane field component large enough to significantly alter the in-plane magnetoresistance measurements [13–17], making it challenging to distinguish the in-plane anisotropy from the field misalignment effect.

The iron-based superconductor represents a unique class of high-temperature superconductors, where the multiorbital character leads to an intriguing interplay between superconductivity, electronic nematic phase, and magnetic order [18–20]. Notably, twofold rotational symmetry has been reported in iron-based superconductors as an experimental signature of the nematic phase [3,4,21,22]. As an archetypal iron-based superconductor in the 2D limit, $FeSe/SrTiO_3$ (FeSe/STO) has attracted considerable interest due to the dramatic enhancement of the superconducting transition temperature and a large superconducting gap compared with bulk FeSe [23–29]. Highly anisotropic upper critical fields of FeSe/STO render $R(\varphi)$ measurements particularly sensitive to field misalignment, and a previous study ascribed the anisotropic in-plane magnetoresistance in FeSe/STO to an unavoidable out-of-plane component of the in-plane magnetic field [13].

In this work, we systematically investigate the angular dependence of the in-plane magnetoresistance $R(\varphi)$ in molecular beam epitaxy (MBE)-grown ultrathin FeSe films on insulating STO substrates. We demonstrate that two distinct $R(\varphi)$ line shapes can be obtained by simply remounting the sample, indicating a dominant contribution of the field misalignment effect. We then develop a geometric model that accounts for the misalignment between the sample plane, the rotation plane, and the magnetic field, which can well reproduce the experimental anisotropic in-plane magnetoresistance in the superconducting transition region of FeSe/STO.

## II. EXPERIMENTAL METHODS

Ultrathin FeSe films with thicknesses of 1–3 unit cells (UCs) were epitaxially grown on undoped insulating STO (001) substrates in a Scienta Omicron MBE system with a base pressure lower than $2.0 \times 10^{-10}$ mbar. The STO substrates were prepared via standard chemical etching and annealing procedures to obtain atomically flat $TiO_2$-terminated surfaces. Subsequently, the substrates were transferred into the MBE chamber and degassed at 380 °C for 1 h and annealed at 480 °C for 1 h. During the growth process, the substrates were maintained at 395 °C. FeSe films were grown by coevaporating high-purity Fe (99.995%) and Se (99.9999%) from standard Knudsen cells with a flux ratio of about 1:20. The growth rate was approximately 0.25 UC/min. After deposition, FeSe films were annealed at 450 °C for several hours to optimize the superconducting transition temperature. For *ex situ* transport measurements, 10-UC FeTe protection layers were capped on top of the FeSe film by coevaporating Fe (99.995%) and Te (99.9999%) with a flux ratio of about 1:10. The transport measurements were conducted using a standard four-probe method in a 14 T Physical Property Measurement System (Quantum Design, PPMS-14) with a rotator option.

## III. RESULTS AND DISCUSSION

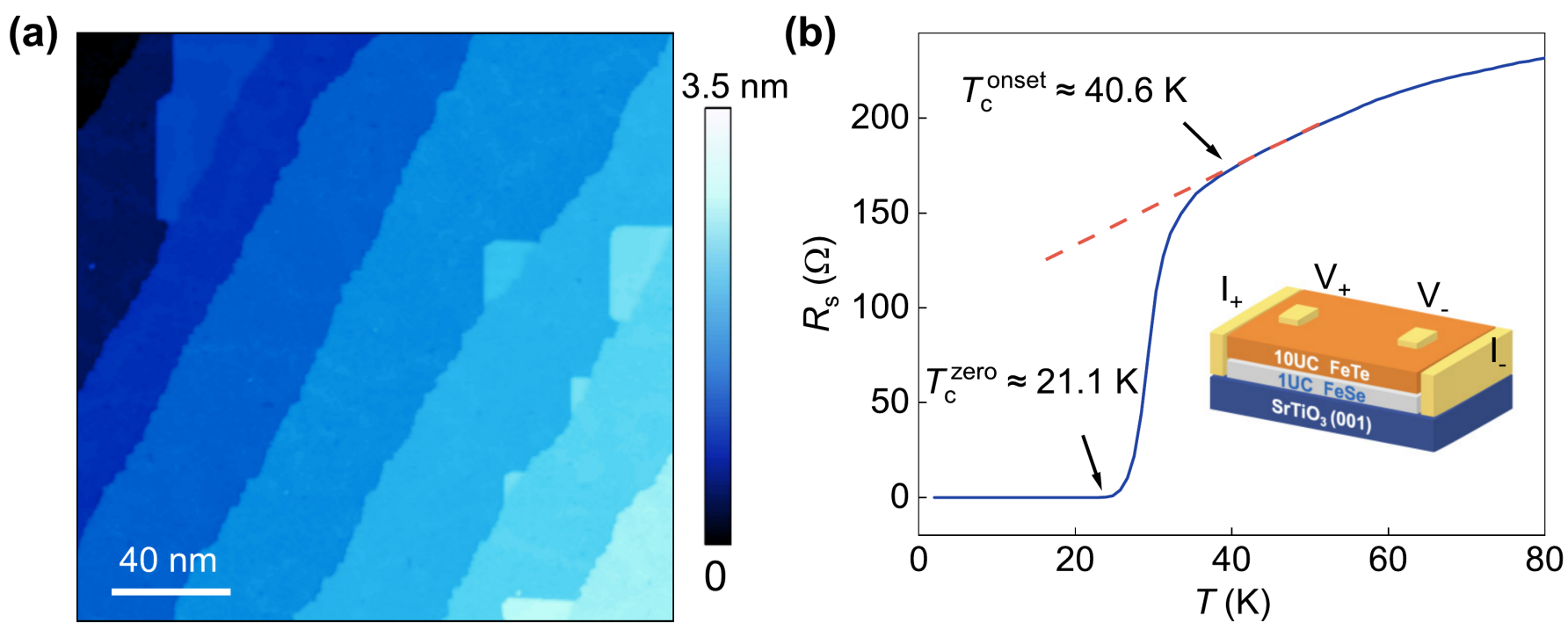

FIG. 1. STM topography and superconducting property of FeSe film on STO substrate. (a) STM topography (200 nm × 200 nm, sample bias $V_s$ = 1 V, tunneling current $I_t$ = 200 pA) of FeSe film on Nb-doped STO. (b) Temperature dependence of sheet resistance $R_s(T)$ of FeSe film on insulating STO at zero field, showing $T_c^{onset} \approx 40.6$ K and $T_c^{zero} \approx 21.1$ K (sample 1). The inset is a schematic for standard four-electrode transport measurements.

Figure 1(a) displays the scanning tunneling microscopy (STM) topographic image of a typical FeSe film epitaxially grown on a STO substrate, revealing atomically flat terraces originating from the STO substrate. After capping the 10-UC FeTe protection layer, we performed *ex situ* transport measurements on a representative FeSe film (sample 1) using the standard four-probe method [see inset of Fig. 1(b) for a schematic]. The temperature dependence of the sheet resistance is shown in Fig. 1(b), exhibiting a superconducting transition with an onset temperature $T_c^{onset} \approx 40.6$ K and zero resistance within the measurement resolution at $T_c^{zero} \approx 21.1$ K. Here, the onset temperature is defined as the temperature below which $R_s$ deviates from the linear background of the normal state. Superconducting properties of samples 1–4 are displayed in Fig. S2 in the Supplemental Material (SM) [30]. It is worth noting that the superconductivity in FeSe/STO is localized in the first unit cell of FeSe at the interface as evidenced by previous scanning tunneling microscopy studies [23,31].

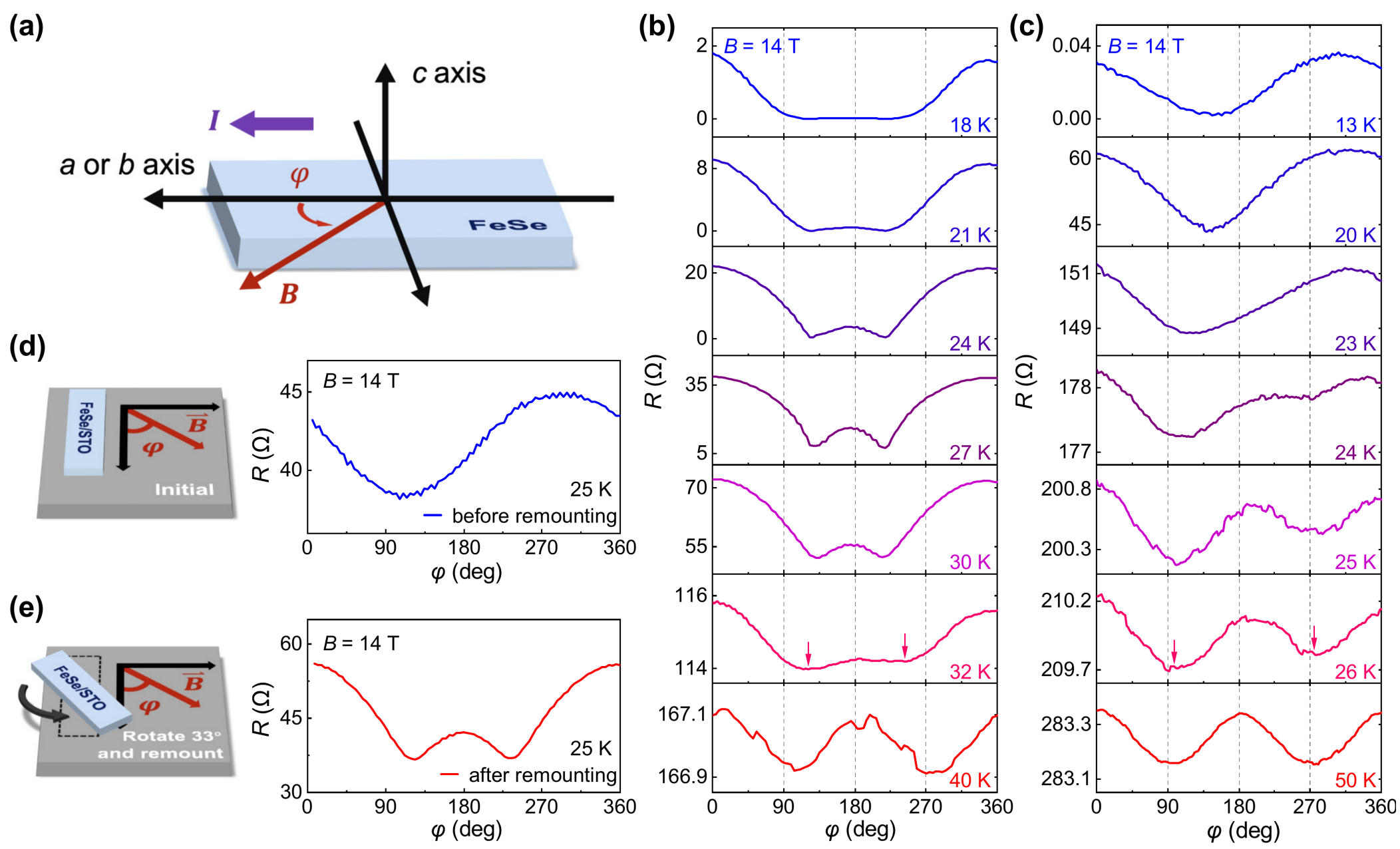

FIG. 2. Angular dependence of in-plane magnetoresistance in FeSe/STO. (a) Schematic for the angular dependent in-plane magnetoresistance measurements. Here, $\varphi$ represents the angle between the magnetic field (red arrow) and the *a* or *b* axis of the FeSe sample. The excitation current (purple arrow) is applied along the *a* or *b* axis. (b), (c) The angular dependence of in-plane magnetoresistance $R(\varphi)$ measured at various temperatures under $B = 14$ T for sample 1 and sample 2, respectively. The pink arrows indicate the asymmetric double-dip feature. (d), (e) Angular dependent in-plane magnetoresistance measurements for sample 3 at $B = 14$ T before (d) and after (e) remounting the sample with an in-plane rotation of 33°. The left panels show schematics of the configurations before and after remounting.

To study the in-plane anisotropy of FeSe/STO, we measured the angular dependence of the in-plane magnetoresistance $R(\varphi)$ using a sample rotator. Here, $\varphi$ is defined as the angle between the magnetic field and the *a* or *b* axis of the FeSe film, and the excitation current is applied along the *a* or *b* axis [see Fig. 2(a) for a schematic]. Figures 2(b) and 2(c) present $R(\varphi)$ curves of two representative FeSe films (samples 1 and 2) at different temperatures under 14 T. In the high-temperature regime (around or above $T_c^{onset}$), the $R(\varphi)$ curves of both samples exhibit twofold rotational symmetry. At low temperatures, the $R(\varphi)$ curve of sample 1 evolves into a symmetric double-dip feature (showing the same resistance at two dips), whereas the $R(\varphi)$ curve of sample 2 develops a cosinusoidal-like single-dip feature [Fig. 2(c)].

To investigate the origin of the distinct $R(\varphi)$ behaviors, we performed a control experiment on sample 3. After carefully remounting the sample with an in-plane rotation of approximately 33° [left panels of Figs. 2(d) and 2(e)], the $R(\varphi)$ curve at 25 K transforms from a single-dip feature into a double-dip feature [Figs. 2(d) and 2(e)]. Note that the $R_s(T)$ curve of sample 3 remains essentially unchanged after remounting (Fig. S3 in SM [30]), indicating the remounting procedure does not change the superconducting properties. Therefore, the single-dip and double-dip features observed in $R(\varphi)$ curves at low temperatures originate from the magnetic field misalignment related to sample mounting rather than the intrinsic property of FeSe film.

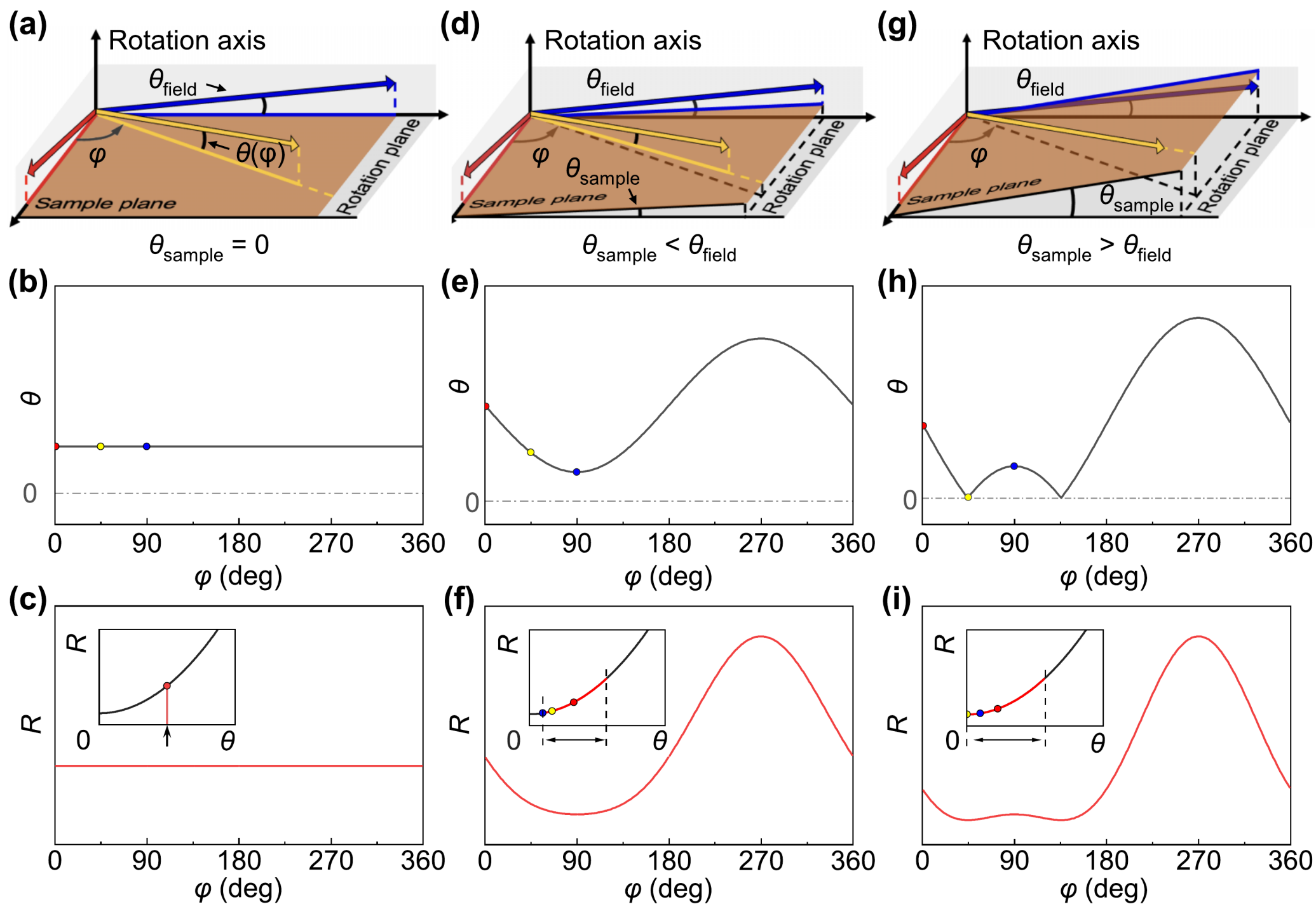

FIG. 3. Geometric analysis of field misalignment and its effect on the anisotropic in-plane magnetoresistance $R(\varphi)$. (a) Geometry of a simple case: the sample plane coincides with the rotation plane $\theta_{\text{sample}} = 0°$, and the magnetic field has a tilt angle relative to the rotation plane $\theta_{\text{field}} > 0°$. Here, $\varphi$ denotes the rotation angle of the magnetic field, and $\theta(\varphi)$ is the angle between the magnetic field vector and the sample plane. The colored arrows indicate field directions at $\varphi = 0°$ (red), 45° (yellow) and 90° (blue), and the colored lines are projections of the field vector on the sample plane. (b) Calculated $\theta(\varphi)$ for case (a). The colored solid circles mark the values corresponding to the colored arrows in (a). (c) Simulated $R(\varphi)$ for case (a). The inset shows the assumed polar angular dependence of the magnetoresistance $R(\theta)$. (d)–(f) Same as (a)–(c), but for the case where both sample and field are tilted relative to the rotation plane and $\theta_{\text{sample}} < \theta_{\text{field}}$. (g)–(i) Same as (a)–(c), but for the case where $\theta_{\text{sample}} > \theta_{\text{field}}$. We note that the double-dip feature shown in Figs. 2(b) and 2(e) exhibits a phase shift of approximately 90° relative to the feature in Fig. 3(i), which is attributed to a different initial phase $\varphi_0$ associated with the sample mounting.

Given the strong anisotropy between out-of-plane and in-plane critical fields of FeSe/STO [24], even a small out-of-plane component of a magnetic field could significantly alter the behavior of anisotropic in-plane magnetoresistance. Thus we model the experimental observations by considering a small out-of-plane component of the magnetic field and a small tilt between the sample plane and the rotation plane. As a result of field misalignment, the out-of-plane angle $\theta$ of the magnetic field relative to the sample plane varies as a function of the azimuthal angle $\varphi$. Using the coordinate definitions in Fig. 3(a) and Fig. S1 in SM [30], the effective out-of-plane angle $\theta(\varphi)$ can be obtained from the projection of the rotating field vector onto the sample-plane normal (see SM [30] for details):

$$\theta(\varphi) = \arcsin\left[\left|\sin(\theta_{\text{field}})\cos(\theta_{\text{sample}}) - \cos(\theta_{\text{field}})\sin(\theta_{\text{sample}})\sin(\varphi - \varphi_0)\right|\right] \quad (1)$$

where $\varphi_0$ is a fixed angle that specifies the sample plane relative to the rotator frame for a given sample mounting (see Fig. S1 in SM [30] for details), $\theta_{\text{field}}$ and $\theta_{\text{sample}}$ denote the tilt angles of the magnetic field and the sample plane relative to the rotation plane, respectively. To build the connection between the anisotropic in-plane resistance $R(\varphi)$ and $\theta(\varphi)$, we assume that the magnetoresistance follows a quadratic dependence on $\theta$ [the inset of Fig. 3(c)]. For a given $\varphi$, we calculate $\theta(\varphi)$ from Eq. (1) and then evaluate the resistance from the $R(\theta)$ curve. Figure 3(a) illustrates a simple case where the sample plane coincides with the rotation plane ($\theta_{\text{sample}} = 0°$) and the magnetic field has a tilt angle relative to the rotation plane ($\theta_{\text{field}} > 0°$). In this case, Eq. (1) simplifies to $\theta(\varphi) = \theta_{\text{field}}$, and $R(\varphi)$ is a constant [Figs. 3(b) and 3(c)]. For the case in Fig. 3(d), both sample and field are tilted relative to the rotation plane and $\theta_{\text{sample}} < \theta_{\text{field}}$. The magnetic field vector remains on the same side of the sample plane during rotation, giving rise to a cosinusoidal-like dependence in $\theta(\varphi)$ [Fig. 3(e)] and further results in a single-dip feature in the $R(\varphi)$ curve [Fig. 3(f)]. For the case $\theta_{\text{sample}} > \theta_{\text{field}}$ [Fig. 3(g)], the magnetic field vector crosses the sample plane [$\theta(\varphi) = 0°$] twice when $\varphi$ rotates from 0° to 360°. Consequently, the $\theta(\varphi)$ curve shows two minima [Fig. 3(h)], leading to a symmetrical double-dip feature in $R(\varphi)$ [Fig. 3(i)].

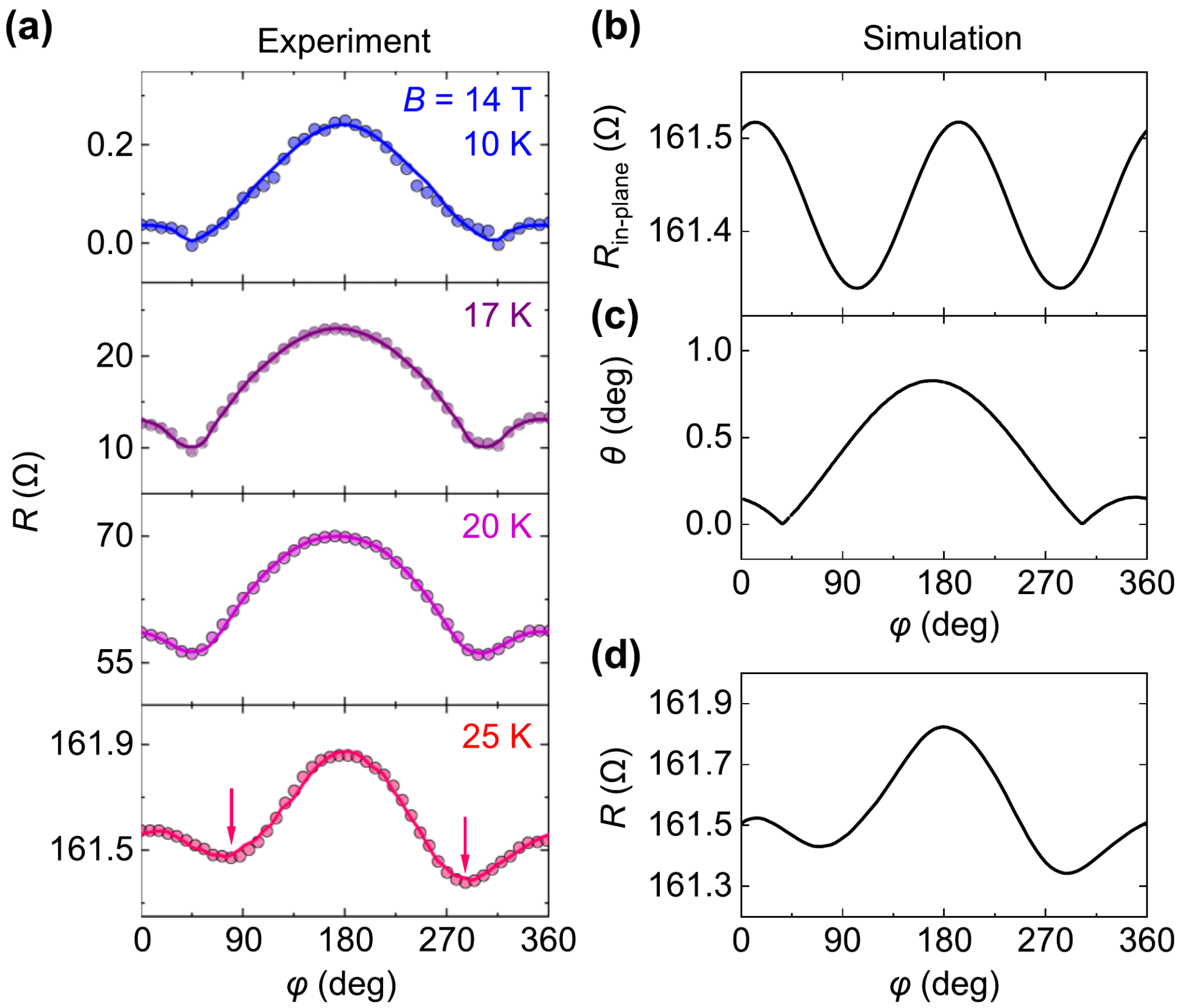

FIG. 4. Anisotropic in-plane magnetoresistance $R(\varphi)$ in FeSe/STO. (a) Measured $R(\varphi)$ for sample 4 at various temperatures under $B$ = 14 T, which can be well fitted by our model considering a twofold in-plane symmetry and the field misalignment effect. The pink arrows indicate the asymmetric double-dip feature. (b) Assumed twofold in-plane anisotropy. (c) Simulated out-of-plane angle $\theta(\varphi)$ relative to the sample plane. The feature here has a phase shift of approximately 90° compared to Fig. 3(h). This phase shift is attributed to a different initial phase $\varphi_0$ caused by sample mounting. (d) Simulated $R(\varphi)$ considering a twofold in-plane symmetry (b) and the field misalignment effect (c).

Moreover, the measured $R(\varphi)$ curve of FeSe/STO often exhibits an asymmetric double-dip feature (showing different resistance at two dips, marked by pink arrows) just below $T_c^{onset}$ [Figs. 2(b), 2(c), and 4(a)], which cannot be explained by field misalignment alone. To account for this asymmetry, we include in our model an intrinsic in-plane magnetoresistance term $R_{in\text{-}plane}(\varphi)$ with a general $N$-fold symmetry form:

$$R_{in-plane}(\varphi) = A\cos(N\varphi + \varphi_1) + R_c \quad (2)$$

Here, $R_c$ is a constant independent of $\varphi$, $A$ is the amplitude, $\varphi_1$ is the initial phase, and $N$ determines the $N$-fold symmetry. In our model, the $R_{\text{in-plane}}(\varphi)$ modulates the in-plane magnetoresistance at $\theta = 0°$, while the $R(\theta)$ term can be estimated using the measured $R(\theta)$ curves (Fig. S4 in SM [30]). After effectively incorporating the twofold in-plane anisotropy of the magnetoresistance and the field misalignment effect, our simulated $R(\varphi)$ curves show excellent agreement with the experimental $R(\varphi)$ curves of sample 4 measured at various temperatures [Fig. 4(a)]. To clarify the origin of the experimental $R(\varphi)$, we break down the simulation processes in Figs. 4(b)–4(d). When taking the twofold in-plane anisotropy [Fig. 4(b)] and the field misalignment effect [Fig. 4(c)] into account, the simulated $R(\varphi)$ curve successfully reproduces the asymmetric double-dip feature, as shown in Fig. 4(d). At lower temperatures ($T \leq 20$ K), the $R(\varphi)$ curves exhibit typical double-dip features [Fig. 4(a)]. The best fits yield $\theta_{\text{field}} = 0.34°$ and $\theta_{\text{sample}} = 0.52°$, consistent with the case $\theta_{\text{sample}} > \theta_{\text{field}}$ [Fig. 3(g)]. Notably, the fitted amplitude of the twofold in-plane anisotropy is vanishingly small when $T \leq 20$ K. Therefore, the observed $R(\varphi)$ features near $T_c^{\text{zero}}$ are attributed primarily to the out-of-plane magnetic field component arising from the misalignment. It is noteworthy that, although our model can disentangle the field misalignment effect from the in-plane magnetoresistance, the underlying mechanism of the twofold in-plane magnetoresistance in FeSe/STO remains not fully understood and requires further investigation.

The field misalignment effect has been a long-standing issue in 2D or layered superconductors with highly anisotropic upper critical fields [13–17], as a small out-of-plane field component can significantly alter the measured in-plane magnetoresistance and may produce anisotropic features that could otherwise be interpreted as rotational symmetry breaking. In previous studies, field misalignment was mainly discussed qualitatively or estimated using a simple projection model, in which the misalignment angle was estimated by projecting the magnetic field along the out-of-plane or in-plane orientation. The field misalignment contribution was ruled out as the dominant origin of the observed anisotropy when the estimated misalignment angle was much larger than the possible angular deviation in those experiments [11,12,15]. A two-axis rotator can significantly improve the alignment in anisotropic in-plane magnetoresistance measurements, and the best in-plane alignment is determined by scanning the polar angle at each azimuthal angle to locate a magnetoresistance minimum [14]. Although this procedure enables precise alignment, it requires repeated scans and is very time consuming, taking approximately 30 times as long as single-axis rotation measurements. In this work, we address the field misalignment issue following a complementary route. By considering the geometric misalignment and the possible in-plane anisotropy, our model quantitatively reproduces the observed single-dip and double-dip $R(\varphi)$ line shapes and identifies the corresponding misalignment conditions. Our work provides a practical procedure for evaluating whether subtle anisotropic features in angle-dependent in-plane magnetoresistance measurements may originate from geometric misalignment, thereby providing useful feedback for calibration and establishing a more reliable basis for identifying intrinsic rotational symmetry breaking. This model is also broadly applicable to in-plane magnetoresistance measurements with different device geometries. In practical measurements, not only is the standard four-probe method highly susceptible to the field misalignment effect, but the Corbino measurements can also be affected by out-of-plane field components (Fig. S5) in SM [30].

## IV. CONCLUSION

In summary, we systematically investigated the azimuthal angular dependence of the in-plane magnetoresistance $R(\varphi)$ in MBE-grown FeSe films on STO substrates. Below $T_c^{onset}$, $R(\varphi)$ develops an asymmetric double-dip feature that eventually evolves into either a single-dip or a double-dip behavior at lower temperatures. Our geometric analysis indicates that these features mainly originate from the small misalignment of magnetic field relative to the sample plane. Our study provides a practical framework for quantitatively assessing and disentangling the field misalignment effect in angular-dependent measurements on 2D or layered superconductors.

## ACKNOWLEDGMENTS

We thank Wei Ren, Rao Li, Longxin Pan, Tiezheng Zhang, Yanan Li, and Guanyang He for their assistance in sample growth. This work was financially supported by the National Key Research and Development Program of China (Grant No. 2022YFA1403103), the National Natural Science Foundation of China (Grant No. 12488201), the Quantum Science and Technology National Science and Technology Major Project (Grant No. 2021ZD0302403), the National Key Research and Development Program of China (Grant No. 2023YFA1406500), the National Natural

Science Foundation of China (Grants No. 12174442, No. 92565303, No. 12404215, and No. 92565305), the Fundamental Research Funds for the Central Universities and the Research Funds of Renmin University of China (Grants No. 22XNKJ20 and No. 26XNKJ22), and the Guangdong Provincial Quantum Science Strategic Initiative (Grants No. GDZX2401001, No. GDZX2401009, No. GDZX2603006 and No. GDZX2601001).

## DATA AVAILABILITY

The data that support the findings of this article are not publicly available. The data are available from the authors upon reasonable request.

Supplemental Material for

# Field misalignment effect on the in-plane magnetoresistance in high-temperature superconductor $FeSe/SrTiO_3$

Jie Liu[1,*], Zihan Cui[1,*], Xiaorong Liu[1], Yanzhao Liu[2], Ziqiao Wang[2,†], Jian Wang[3,4,5,6,‡], and Yi Liu[1,2,7,§]

[1] *School of Physics and Key Laboratory of Quantum State Construction and Manipulation (Ministry of Education), Renmin University of China, Beijing 100872, China*
[2] *Quantum Science Center of Guangdong–Hong Kong–Macao Greater Bay Area (Guangdong), Shenzhen 518045, China*
[3] *International Center for Quantum Materials, School of Physics, Peking University, Beijing 100871, China*
[4] *Beijing Key Laboratory of Quantum Devices, Peking University, Beijing 100871, China*
[5] *Hefei National Laboratory, Hefei 230088, China*
[6] *Collaborative Innovation Center of Quantum Matter, Beijing 100871, China*
[7] *Xiongan Future Industrial Institute, Renmin University of China, Xiongan 071700, China*

[*]These authors contributed equally to this work.
[†]Contact author: wangziqiao@quantumsc.cn
[‡]Contact author: jianwangphysics@pku.edu.cn
[§]Contact author: yiliu@ruc.edu.cn

**Derivation of the Effective Tilt Angle $\theta(\varphi)$**

For ideal angular dependent in-plane magnetoresistance measurements $R(\varphi)$, the applied magnetic field $\vec{\boldsymbol{B}}$ is strictly parallel to the sample plane and rotates within that plane as the azimuthal angle $\varphi$ changes. However, in practice, small field misalignments are unavoidable, which originates from a small tilt between the sample plane and the rotation plane and a small out-of-plane component of magnetic field. Consequently, the out-of-plane angle $\theta$ of $\vec{\boldsymbol{B}}$ relative to sample plane changes with respect to $\varphi$, which further alter the behavior of the measured $R(\varphi)$. To resolve the field misalignment effect on the angular dependent in-plane magnetoresistance, we analyze the geometric relation and derive an expression for the effective out-of-plane angle $\theta(\varphi)$.

Figure S1 illustrates the schematic of geometric relation between the rotation plane, the sample plane and the magnetic field in uniaxial rotation measurements. The rotator frame is defined as ($x$, $y$, $z$), where the $z$ axis is aligned with the rotation axis. The magnetic field $\vec{\boldsymbol{B}}$ rotates in the $xy$ plane, with the azimuthal angle $\varphi$ defined in Fig. S1(c). Due to the small tilt angle between the sample plane and the rotation plane, the out-of-plane angle $\theta$ of magnetic field relative to sample plane is a function of $\varphi$. In our measurements, $\theta(\varphi)$ is a small positive angle within 2°.

The sample frame ($x'$, $y'$, $z'$) is defined in Fig. S1(a) and (b). Here, the $z'$ axis is normal to the sample plane, and the $x'$ axis aligns with the intersection line between the sample and rotation planes. Under this configuration, the relative orientation between the sample and rotation planes is determined by two angles. The first is the tilt angle $\theta_{\text{sample}}$, which denotes the inclination between the sample and rotation planes. In the ($x$, $y$, $z$) frame, this tilt corresponds to a rotation about the $x$ axis, represented by the transformation matrix:

$$\boldsymbol{\Omega}_x = \begin{bmatrix} 1 & 0 & 0 \\ 0 & \cos\theta_{\text{sample}} & -\sin\theta_{\text{sample}} \\ 0 & \sin\theta_{\text{sample}} & \cos\theta_{\text{sample}} \end{bmatrix} \quad (1)$$

The second is the azimuthal angle $\varphi_0$, defined as the angle between the $x'$ axis and the $x$ axis (Fig. S1(b)). For a given sample mounting, $\varphi_0$ is a fixed constant. In the ($x$, $y$,

$z$) frame, this angle corresponds to a rotation about the $z$ axis, represented by the transformation matrix:

$$\boldsymbol{\Omega}_{\mathrm{z}} = \begin{bmatrix} \cos\varphi_0 & -\sin\varphi_0 & 0 \\ \sin\varphi_0 & \cos\varphi_0 & 0 \\ 0 & 0 & 1 \end{bmatrix} \tag{2}$$

Since both steps are extrinsic rotations with respect to the ($x$, $y$, $z$) frame, the total transformation matrix $\boldsymbol{\Omega}_{\mathrm{total}}$ is obtained by:

$$\begin{aligned} \boldsymbol{\Omega}_{\mathrm{total}} &= \boldsymbol{\Omega}_z \times \boldsymbol{\Omega}_x \\ &= \begin{bmatrix} \cos\varphi_0 & -\cos\theta_{\mathrm{sample}}\sin\varphi_0 & \sin\theta_{\mathrm{sample}}\sin\varphi_0 \\ \sin\varphi_0 & \cos\theta_{\mathrm{sample}}\cos\varphi_0 & -\sin\theta_{\mathrm{sample}}\cos\varphi_0 \\ 0 & \sin\theta_{\mathrm{sample}} & \cos\theta_{\mathrm{sample}} \end{bmatrix} \end{aligned} \tag{3}$$

Note that the normal vector $\vec{\boldsymbol{n}}_{\boldsymbol{sample}}$ of sample plane corresponds to the $z'$ axis. Thus, it can be obtained by applying the total transformation matrix $\boldsymbol{\Omega}_{\mathrm{total}}$ to the initial $z$ axis unit vector $\vec{\boldsymbol{e}}_{\boldsymbol{z}}$ = (0, 0, 1)$^{\mathrm{T}}$. The resulting vector is the third column of $\boldsymbol{\Omega}_{\mathrm{total}}$:

$$\vec{\boldsymbol{n}}_{\mathrm{sample}} = \begin{bmatrix} \sin\theta_{\mathrm{sample}}\sin\varphi_0 \\ -\sin\theta_{\mathrm{sample}}\cos\varphi_0 \\ \cos\theta_{\mathrm{sample}} \end{bmatrix} \tag{4}$$

To determine the effective tilt angle $\theta(\varphi)$, we express the magnetic field vector $\vec{\boldsymbol{B}}$ in the ($x$, $y$, $z$) frame. Due to the small misalignment between the magnetic field and the rotation plane, the field maintains a constant tilt angle $\theta_{\mathrm{field}}$ relative to the rotation plane (Fig. S1(c)). For any given azimuthal $\varphi$, the vector can be given by:

$$\vec{\boldsymbol{B}}(\varphi) = B\begin{bmatrix} \cos\theta_{\mathrm{field}}\cos\varphi \\ \cos\theta_{\mathrm{field}}\sin\varphi \\ \sin\theta_{\mathrm{field}} \end{bmatrix} \tag{5}$$

Thus, $\theta(\varphi)$ can be calculated by projecting $\vec{\boldsymbol{B}}$ onto the normal vector of the sample plane ($\vec{\boldsymbol{n}}_{\boldsymbol{sample}}$) for any given rotation angle $\varphi$:

$$\sin\theta(\varphi) = \frac{\left|\vec{\boldsymbol{B}}(\varphi)\cdot\vec{\boldsymbol{n}}_{\mathrm{sample}}\right|}{\left|\vec{\boldsymbol{B}}(\varphi)\right|} \tag{6}$$

By substituting the expressions from Eqs. (4-6), we obtain the final expression of $\theta(\varphi)$:

$$\begin{aligned} \theta(\varphi) &= \arcsin\left[\,\left|\vec{\boldsymbol{B}}(\varphi)\cdot\vec{\boldsymbol{n}}_{\mathrm{sample}}\right|/\left|\vec{\boldsymbol{B}}(\varphi)\right|\,\right] \\ &= \arcsin\left[\,\left|\sin\theta_{\mathrm{field}}\cos\theta_{\mathrm{sample}} - \cos\theta_{\mathrm{field}}\sin\theta_{\mathrm{sample}}\left(\sin\varphi\cos\varphi_0 - \cos\varphi\sin\varphi_0\right)\right|\,\right] \\ &= \arcsin\left\{\,\left|\sin\theta_{\mathrm{field}}\cos\theta_{\mathrm{sample}} - \cos\theta_{\mathrm{field}}\sin\theta_{\mathrm{sample}}\left[\sin(\varphi-\varphi_0)\right]\right|\,\right\} \end{aligned} \tag{7}$$

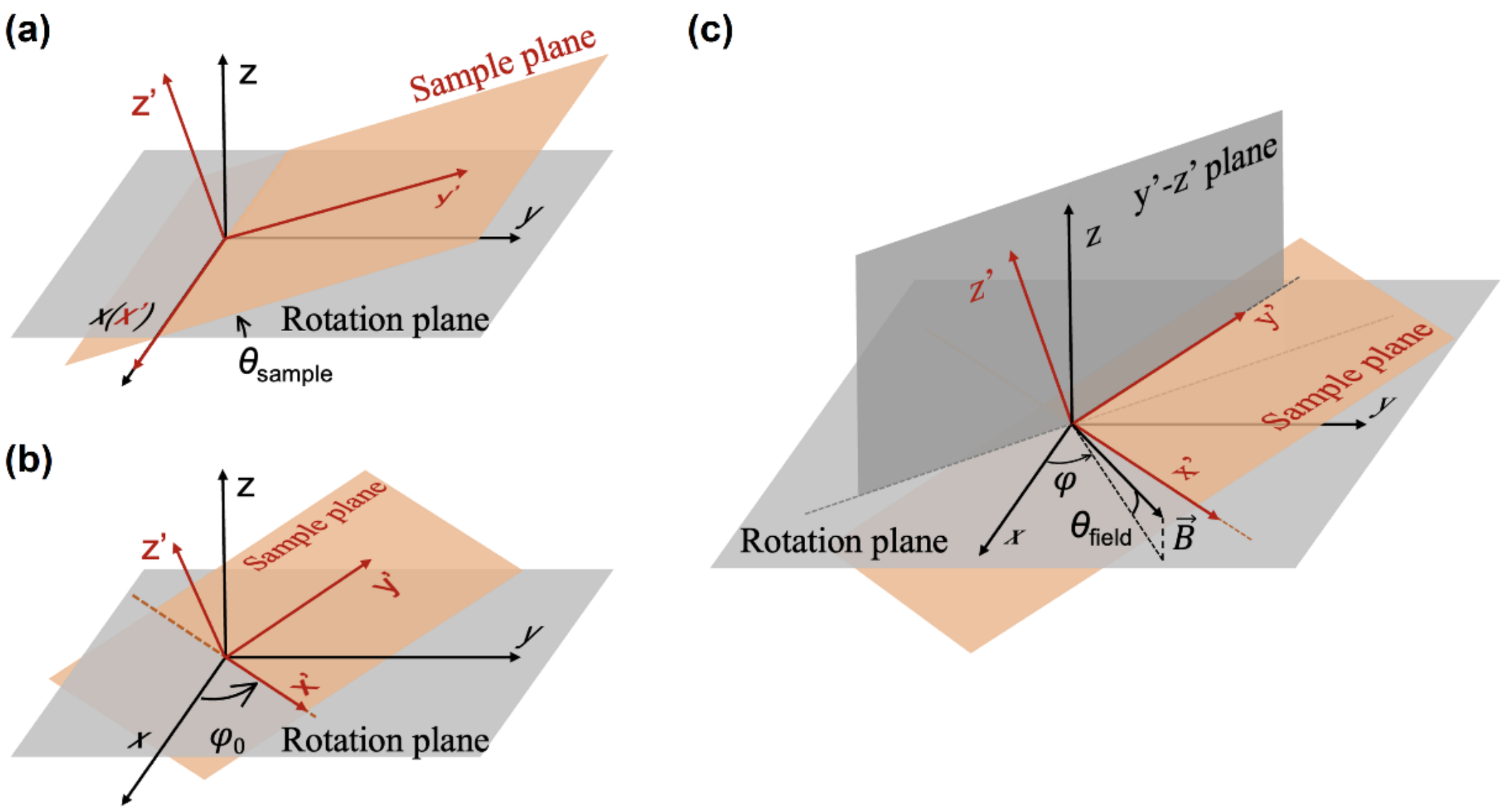

**Fig. S1. Schematic of the geometric relation between the rotation plane, the sample plane and the magnetic field in uniaxial rotation measurements.** In the rotator frame ($x$, $y$, $z$), $z$ axis is the rotation axis and the $xy$ plane is the rotation plane; in the sample frame ($x'$, $y'$, $z'$), $z'$ axis is along the sample normal and the $x'y'$ plane is the sample plane. (a) The tilt of sample plane relative to the rotation plane is described by $\theta_{\mathrm{sample}}$, with the $x'$ axis aligned along the intersection line between the sample plane and the rotation plane. (b) The azimuthal angle $\varphi_0$ is defined as the angle of the $x'$ axis relative to the $x$ axis. (c) The magnetic field vector $\vec{\boldsymbol{B}}$ rotates with an azimuthal angle $\varphi$ and maintains a constant tilt angle $\theta_{\mathrm{field}}$ relative to the rotation plane.

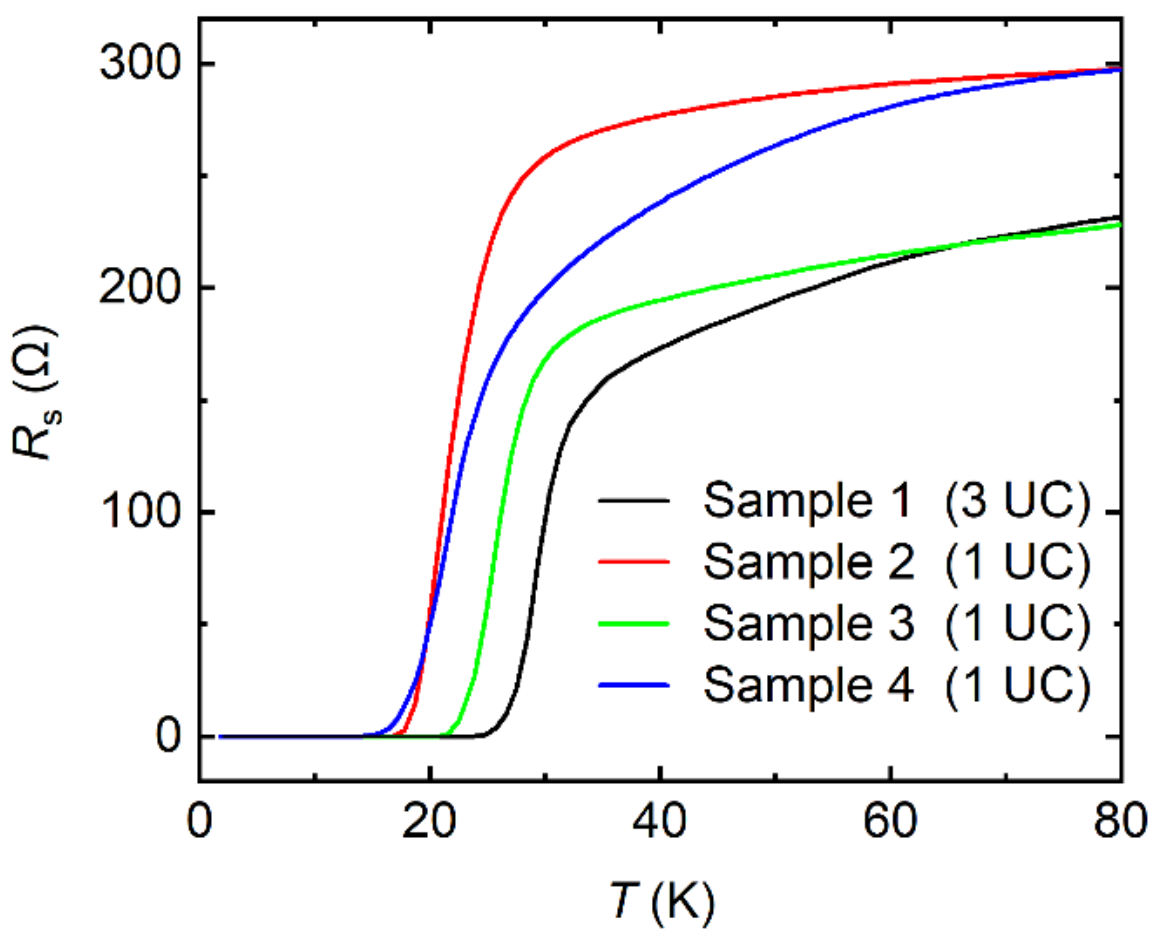

**Fig. S2. Superconducting properties of four representative FeSe/$SrTiO_3$ samples.**

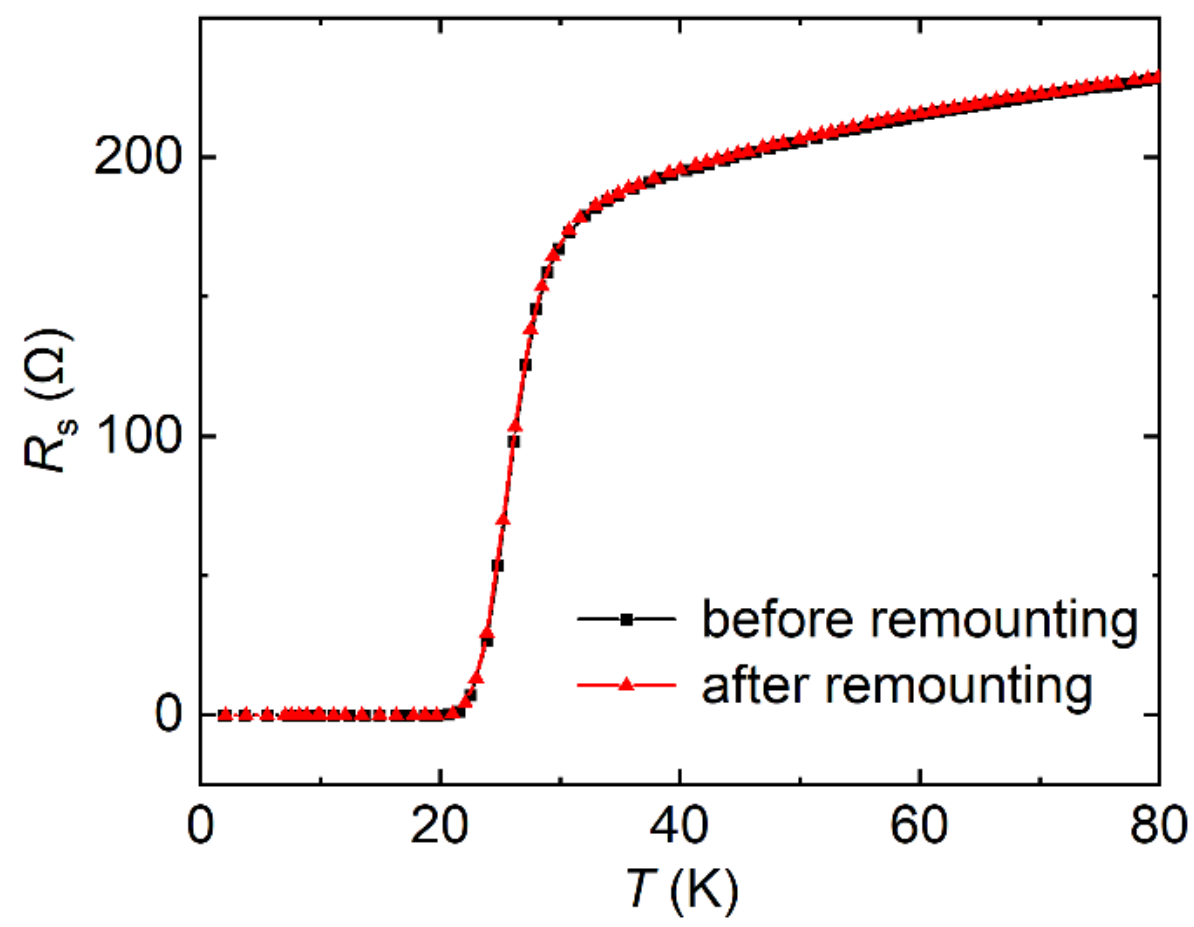

**Fig. S3. Comparison of superconducting property for Sample 3 before and after remounting.** The measured $R_s(T)$ curves measured before and after remounting nearly overlap, indicating that the remounting procedure does not change the superconducting properties.

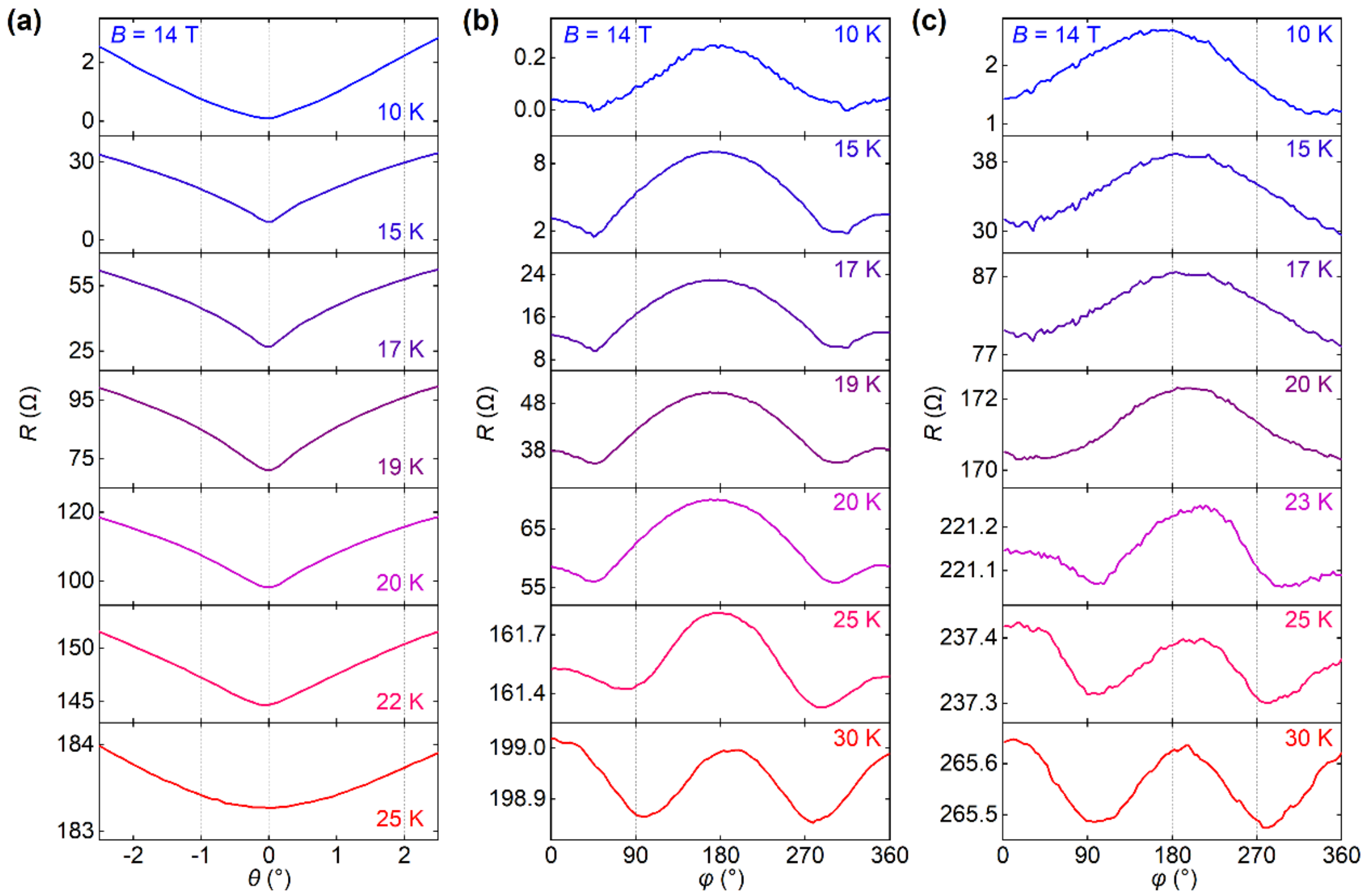

**Fig. S4. Out-of-plane and in-plane anisotropic magnetoresistance for Sample 4 under magnetic field of 14 T at different temperatures.** (a) Polar angular dependence of magnetoresistance $R(\theta)$ at various temperatures. Here, the sign of $\theta$ distinguishes the tilt direction relative to the sample plane. (b) Angular dependence of in-plane magnetoresistance $R(\varphi)$ at various temperatures. (c) same as (b), but measured after remounting.

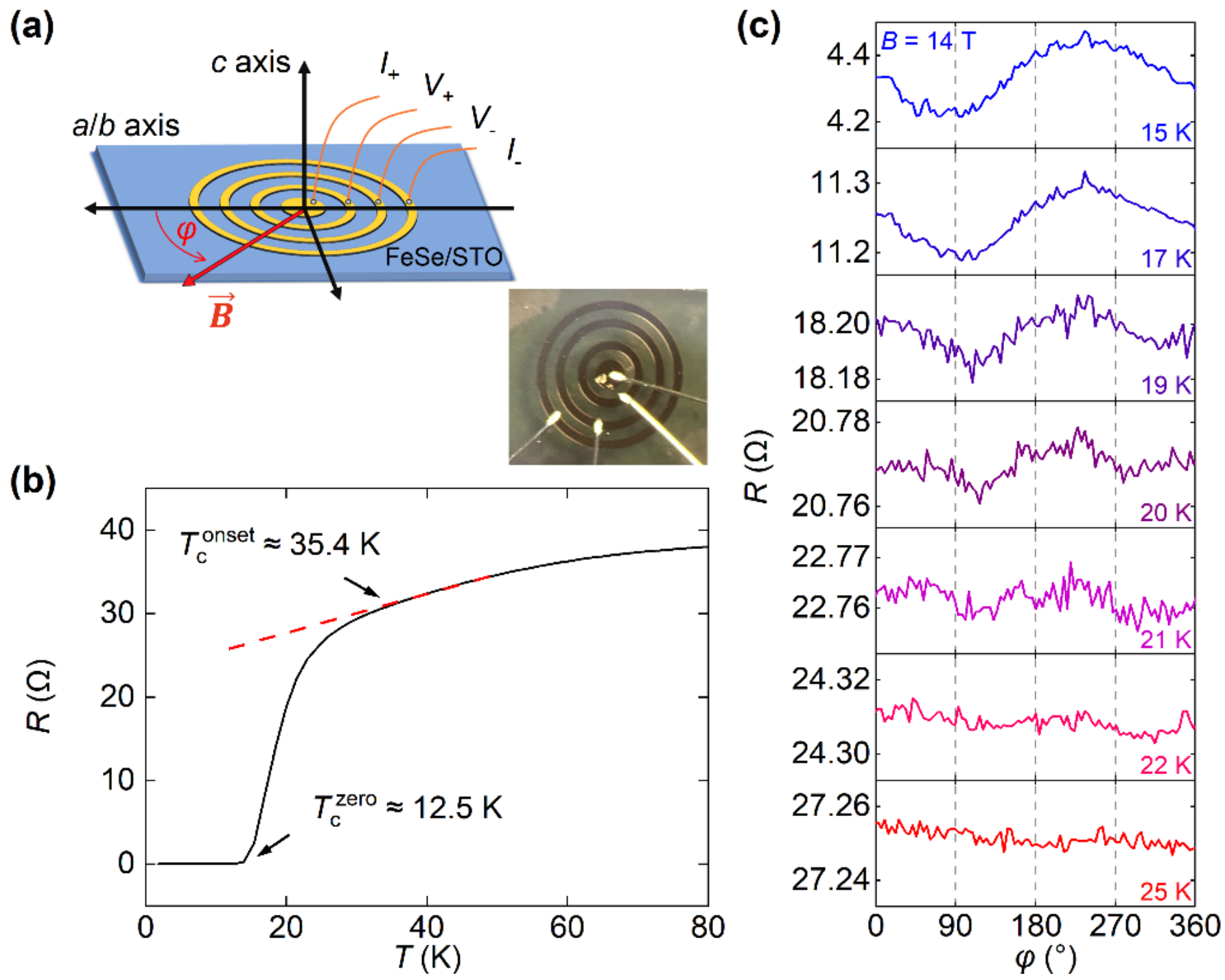

**Fig. S5. Angular dependent in-plane magnetoresistance measurements using the Corbino geometry.** (a) Schematic and optical photo (inset) of the Corbino configuration for angular dependent in-plane magnetoresistance $R(\varphi)$ measurements of the FeSe/STO film. Here, $\varphi$ is defined as the angle between the in-plane magnetic field $\vec{B}$ and the *a*/*b* axis of FeSe/STO. The outer radius of the Corbino electrode is 1 mm. (b) Temperature dependence of resistance $R(T)$ measured in the corbino configuration at zero magnetic field from 2 K to 80 K, which shows $T_c^{onset} \approx 35.4$ K and $T_c^{zero} \approx 12.5$ K. (c) Angular dependence of in-plane magnetoresistance $R(\varphi)$ at various temperatures. Field misalignment effect can also affect the anisotropic in-plane magnetoresistance for Corbino geometry at low temperatures.